\documentclass{article} 
\usepackage{nips15submit_e,times}
\usepackage{hyperref}
\usepackage{url}
\usepackage{amsmath,cite,graphicx,datetime}

\title{Neurons as an Information-theoretic Engine}

\author{
Hideaki Shimazaki \thanks{Webpage \href{http://2000.jukuin.keio.ac.jp/shimazaki}{http://2000.jukuin.keio.ac.jp/shimazaki}, an alternative email address \href{mailto:shimazaki@jhu.edu}{shimazaki@jhu.edu} } \\
RIKEN Brain Science Institute\\
Wako-shi, Satama, Japan\\
\texttt{shimazaki@brain.riken.jp} \\
}

\nipsfinalcopy 

\begin{document}
\begin{flushright} \usdate \today \end{flushright}

\maketitle


\begin{abstract}
We show that dynamical gain modulation of neurons' stimulus response is described as an information-theoretic cycle that generates entropy associated with the stimulus-related activity from entropy produced by the modulation. To articulate this theory, we describe stimulus-evoked activity of a neural population based on the maximum entropy principle with constraints on two types of overlapping activities, one that is controlled by stimulus conditions and the other, termed internal activity, that is regulated internally in an organism. We demonstrate that modulation of the internal activity realises gain control of stimulus response, and controls stimulus information. A cycle of neural dynamics is then introduced to model information processing by the neurons during which the stimulus information is dynamically enhanced by the internal gain-modulation mechanism. Based on the conservation law for entropy production, we demonstrate that the cycle generates entropy ascribed to the stimulus-related activity using entropy supplied by the internal mechanism, analogously to a heat engine that produces work from heat. We provide an efficient cycle that achieves the highest entropic efficiency to retain the stimulus information. The theory allows us to quantify efficiency of the internal computation and its theoretical limit.
\end{abstract}

\section{Introduction}
Humans and animals change sensitivity to sensory stimulus either adaptively to the stimulus conditions or following a behavioural context even if the stimulus does not change. A potential neurophysiological basis underlying these observations is gain modulation that changes responsiveness of neurons to stimulus; an example is contrast gain-control found in retina \cite{Sakmann1969} and primary visual cortex under anaesthesia \cite{Ohzawa1985,Laughlin1989}, or in higher visual area caused by attention \cite{Reynolds2000,MartnezTrujillo2002}. Theoretical considerations suggested the gain modulation as a nonlinear operation that integrates information from different origins, offering ubiquitous computation performed in neural systems (see \cite{Salinas2001,Carandini2012} for reviews). Regulation of the level of background synaptic inputs \cite{Chance2002,Burkitt2003}, shunting inhibition \cite{Doiron2001,Prescott2003, Mitchell2003}, and synaptic depression \cite{Abbott1997,Rothman2009} among others have been suggested as potential biophysical mechanisms of the gain modulation (see \cite{Silver2010} for a review). While such modulation of the informative neural activity is a hallmark of computation performed internally in an organism, a principled view to quantify the internal computation has not been proposed yet. 


Neurons convey information about the stimulus in their activity patterns. To describe probabilities of a combinatorially large number of activity patterns of the neurons with a smaller number of activity features, the maximum entropy principle has been successfully used \cite{Schneidman2006,Shlens2006}. This principle constructs the least structured probability distribution given the small set of specified constraints on the distribution, known as a maximum entropy model. It explains probabilities of activity patterns as a result of nonlinear operation on the specified features using a softmax function. Moreover, the model belongs to an exponential family distribution, or a Gibbs distribution. Equivalence of inference under the maximum entropy principle with aspects of the statistical mechanics and thermodynamics was explicated through the work by E. Jaynes\cite{Jaynes1957}. Recently thermodynamic quantities were used to assess criticality of neural activity\cite{Tkaik2014,Tkaik2015}. However, analysis of neural populations under this framework only recently started to include `dynamics' of a neural population \cite{Shimazaki2009,Shimazaki2012,Shimazaki2013,Kass2011,Kelly2012,GranotAtedgi2013,Nasser2013}, and has not yet reached maturity to include computation performed internally in an organism. 

Based on a neural population model obtained under the maximum entropy principle, this study investigates neural dynamics during which gain of neural response to a stimulus is modulated with a delay by an internal mechanism to enhance the stimulus information. This process is expected for dynamics of neurons subject to a feedback gain-modulation mechanism, e.g., via recurrent networks \cite{Salinas1996,Spratling2004,Sutherland2009}. Regardless of the mechanisms, the delay is observed in the gain modulation at different stages of visual pathways \cite{McAdams1999,Reynolds2000,Lee2003}. For example, effect of contrast gain-control by attention on response of V4 neurons to high contrast stimulus appears 200-300 ms after the stimulus presentation, but is absent during 100-200 ms time period during which the neural response is returning to a spontaneous rate \cite{Reynolds2000}. We demonstrate that our hypothetical dynamics of delayed gain-modulation forms an information-theoretic cycle that generates entropy ascribed to the stimulus-related activity using entropy supplied by the internal gain-modulation mechanism. The process works analogously to a heat engine that produces work from heat supplied by reservoirs. We define entropic efficiency of gain-modulation performed to retain the stimulus information, and provide a cycle that achieves the highest entropic efficiency.

This paper is organised as follows. In Section \ref{sec:MaxEnt} we construct a maximum entropy model of a neural population by constraining two types of activities, one that is directly regulated by stimulus and the other that represents background activity of neurons, termed `internal activity'. We point out that modulation of the internal activity realises gain-modulation of stimulus response. In Section \ref{sec:first_law}, we explain the conservation of entropy, equation of state for the neural population, and information on stimulus. In Section \ref{sec:info_engine}, we construct cycles of neural dynamics that model stimulus-evoked activity during which the stimulus information is enhanced by the internal mechanism. We show that an ideal cycle introduced in this section achieves the highest efficiency in retaining the stimulus information. Derivations of free energies of the neural population are summarised in Appendix. 

\section{A simple model of gain modulation by a maximum entropy model} \label{sec:MaxEnt}
\textbf{Maximum entropy model of spontaneous neural activity.}
We start by modelling spontaneous activity of $N$ spiking neurons. We represent a state of the $i$th neuron by a binary variable $x_i = (0,1)$ ($i=1\cdots N$). Here silence of the neuron is represented by `0' whereas activity, or a spike, of the neuron is denoted by `1'. The simultaneous activity of the $N$ neurons is represented by a vector of the binary variables, $\mathbf{x} = (x_1,\ldots, x_N)$. The joint probability mass function, $p(\mathbf{x})$, describes the probability of generating the pattern $\mathbf{x}$. There are $2^N$ different patterns. We characterise the combinatorial neural activity with a smaller number of characteristic features $F_i(\mathbf{x})$ ($i=1,\ldots,d$, where $d < 2^N$), based on the maximum entropy principle. Here $F_i(\mathbf{x})$ is the $i$th feature that combines the activity of individual neurons. For example, these features can be the first and second order interactions, $F_i(\mathbf{x})=x_i$ for $i=1,\ldots,N$, and $F_{N+(N-i/2)(i-1)+j-i}(\mathbf{x})= x_i x_j$ for $i<j$. The maximum entropy principle constructs the least structured probability distribution while expected values of these features are specified\cite{Jaynes1957}. By representing expectation by $p(\mathbf{x})$ using a bracket $\langle \cdot \rangle$, these constraints are written as $\langle F_i(\mathbf{x}) \rangle=c_i$ ($i=1,\ldots,d$), where $c_i$ is the specified constant.

Maximisation of a function subject to the equality constraints is formulated by the method of Lagrange multipliers that alternatively maximises the following Lagrange function, 
\begin{align}
\mathcal{L}[p] = - \sum_{\textbf{x}} p(\textbf{x})  \log p(\textbf{x})  
    - a \sum_{\textbf{x}} p(\textbf{x}) 
    - \sum_i b_i \left\{ \sum_{\textbf{x}} p(\textbf{x}) F_i(\mathbf{x}) - c_i \right\}, 
\end{align}
where $a$ and $b_i$ ($i=1,\ldots,d$) are the Lagrange multipliers. The Lagrange function is a functional of the probability mass function. By finding a zero point of its variational derivative, we obtain
\begin{align}
p(\mathbf{x}) \sim 
\exp\left(- \sum_i b_i F_i(\mathbf{x}) \right). 
\label{max_ent_model}
\end{align}
The Lagrange parameters $b_i$ are obtained by simultaneously solving $ \frac{\partial \mathcal{L}}{\partial b_i}= \langle F_i(\mathbf{x}) \rangle-c_i = 0$ for $i=1,\ldots,d$. Many gradient algorithms and approximation methods have been developed to search the parameters. Activities of retinal ganglion cells\cite{Schneidman2006,Shlens2006,Tkaik2014,Tkaik2015}, hippocampal\cite{Shimazaki2015}, and cortical neurons\cite{Tang2008,Yu2008,Shimazaki2012} were successfully characterised using Eq.~\ref{max_ent_model}. In the following, we use a vector notation $\mathbf{b}_0=(b_1,...,b_d)^T$ and $\mathbf{F}(\mathbf{x})=(F_1(\mathbf{x}),\ldots,F_d(\mathbf{x}))^T$. Here $\mathcal{H}_0 \equiv \mathbf{b}_0^T \mathbf{F}(\mathbf{x})$ is a Hamiltonian of the spontaneously active neurons. In statistical mechanics, Eq.~\ref{max_ent_model} is identified as the Boltzmann distribution with an unit thermodynamic \textit{beta}. 

\textbf{Maximum entropy model of evoked neural activity.} 
In this subsection, we model evoked activity of neurons caused by changes in extrinsic stimulus conditions. We define a feature of stimulus-related activity as $X(\mathbf{x})= \mathbf{b}_1^T \mathbf{F}(\mathbf{x})$, where elements of $\mathbf{b}_1$ dictate response properties of each feature in $\mathbf{F}(\mathbf{x})$ to a stimulus. For simplicity, we represent the stimulus-related activity by this single feature, and consider that the evoked activity is characterised by the two features, $\mathcal{H}_0(\mathbf{x})$ and $X(\mathbf{x})$. To model it, we constrain expectation of the internal and stimulus features using $U$ and $X$, respectively. Here we assume that $\mathbf{F}(\mathbf{x})$, $\mathbf{b}_0$, and $\mathbf{b}_1$ are known and fixed. For example, this would model responses of visual neurons when we change contrast of a stimulus while fixing the rest of the stimulus properties. The maximum entropy distribution subject to these constraints is again given by the method of Lagrange multipliers. The Lagrange function is given as 
\begin{align}
\mathcal{L}[p] =& - \sum_{\textbf{x}} p(\textbf{x})  \log p(\textbf{x})   \nonumber \\
    & - a \sum_{\textbf{x}} p(\textbf{x}) 
     - \beta \left\{ \sum_{\textbf{x}} p(\textbf{x}) \mathcal{H}_0(\mathbf{x}) - U \right\}
    + \alpha \left\{ \sum_{\textbf{x}} p(\textbf{x}) X(\textbf{x}) - X \right\}.
\end{align}
Here $a$, $\beta$, and $\alpha$ are the Lagrange parameters. By maximising the functional $\mathcal{L}$ with respect to $p$, we obtain the following maximum entropy model,
\begin{align}
p(\textbf{x}) = \exp[ -\beta \mathcal{H}_0(\textbf{x}) + \alpha X(\textbf{x}) -\psi(\beta,\alpha)], 
\label{Exp_model}
\end{align}
where $\psi(\beta,\alpha)(=1+a)$ is a logarithm of a normalisation term. It is computed as 
\begin{align}
\psi(\beta,\alpha) = \log \sum_{\textbf{x}} e^{-\beta \mathcal{H}_0(\textbf{x}) + \alpha X(\textbf{x}) }.
\end{align}
We call $\psi(\beta,\alpha)$ a log-partition function. The Lagrange multipliers, $\beta$ and $\alpha$, are adjusted such that $\left< \mathcal{H}_0(\mathbf{x}) \right>= U$ and $\left< X(\mathbf{x}) \right>= X$. Eq.~\ref{Exp_model} is a softmax function (generalisation of a logistic function to multinomial outputs) that returns the population output from a linear sum of the features weighted by $-\beta$ and $\alpha$. With this view, we may alternatively regard $\beta$ or $\alpha$ as an input parameter that controls $U$ and $X$. Hereafter we simply call $U$ internal activity, and $X$ stimulus-related activity. Similarly, we call $\beta$ an internal component, and $\alpha$ a stimulus component. We consider that the stimulus component $\alpha$ can be controlled by changing extrinsic stimulus conditions that an experimenter can manipulate. The stimulus component is written as $\alpha(s)$ if it is a function of a scalar stimulus condition $s$, such as stimulus contrast. In contrast, the internal component $\beta$ is not directly controllable by the stimulus conditions. The spontaneous activity is modelled at $\beta=1$ and $\alpha=0$.

\begin{figure}[t]
\begin{center}
\includegraphics[width=1\textwidth]{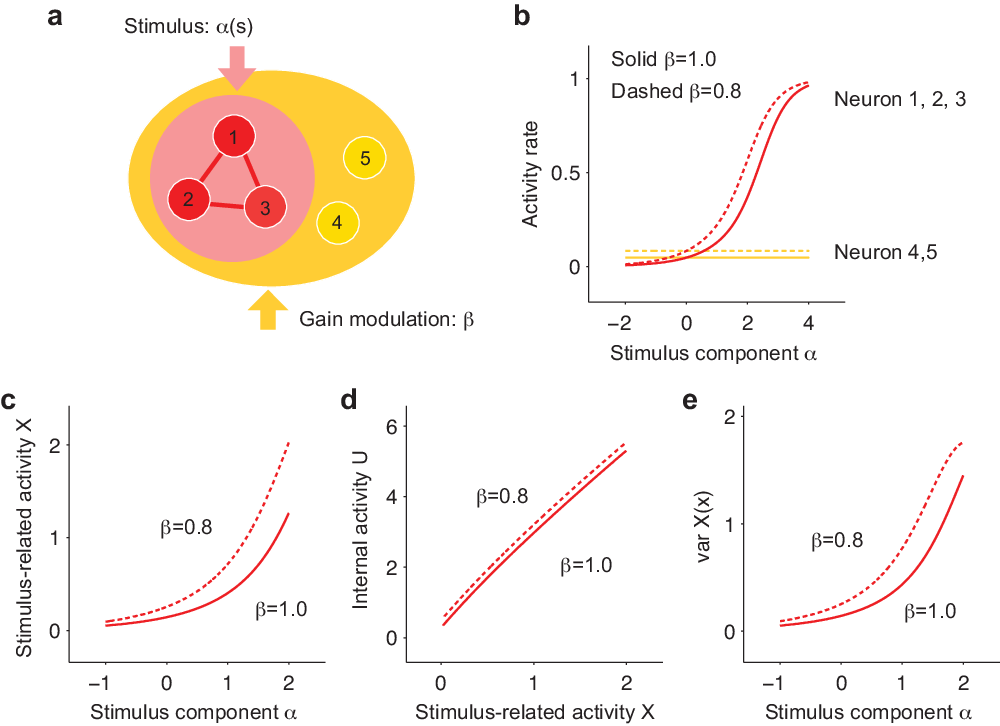}
\end{center}
\caption{
A simple model of gain modulation by a maximum entropy model of 5 neurons. (a) An illustration of neurons that are activated by a stimulus (neurons in a pink area) and controlled by an internal mechanism (neurons in a yellow area). The model is constrained by features containing up to the second order statistics: $\mathbf{F}(\mathbf{x})=(x_1,\ldots,x_5, \ x_1 x_2 ,x_1 x_3,x_2 x_3,\ldots,x_4 x_5)^T$, where the first 5 elements are parameters for the individual activities $x_i$ ($i=1,\ldots,5$) and the rest of the elements is the joint activities of two neurons $x_i x_j$ ($i<j$). We assume that the stimulus-related activity is characterised by $\mathbf{b}_1=(1,1,1,0,0, \ 0.3, 0.3, 0.3, 0, \ldots, 0)$. The first 3 elements are parameters for individual activity of the first three neurons $x_i$ ($i=1,2,3$). The value $0.3$ is assigned to the joint activities of the first three neurons, namely the features specified by $x_1 x_2, x_1 x_3$, and $x_2 x_3$. The internal activity is characterised by $\mathbf{b}_0=(2,2,2,2,2, 0, \ldots, 0)$, which regulates activity rates of individual neurons but does not change their interactions. (b) The activity rates of neurons as a function of the stimulus component $\alpha$ at fixed internal components, $\beta=1.0$ (solid line) and $\beta=0.8$ (dashed line).  (c) The stimulus component $X$ as a function of $\alpha$ at different internal components. (d) The relation between the stimulus-related activity $X$ and internal activity $U$. (e) The Fisher information about the stimulus component $\alpha$. 
}
\label{fig:gain_modulation}
\end{figure}

\textbf{Gain modulation by internal activity.} 
We give a simple example of the maximum entropy model to show how the internal activity modulates the stimulus-related activity. Figure \ref{fig:gain_modulation}a illustrates an exemplary model composed of 5 neurons. With these particular model parameters (see figure caption), the stimulus component $\alpha$ controls activity rates of the first three neurons and their correlations. The internal component $\beta$ controls background activity rates of all neurons. In our settings, decreasing $\beta$ increases the baseline activity level of all neurons. Figure~\ref{fig:gain_modulation}b displays activity rates of the individual neurons ($\langle x_i \rangle$ for $i=1,\ldots,5$) as a function of the stimulus component $\alpha$ with a fixed internal component $\beta$. Increasing $\alpha$ under these conditions activates the first three neurons without changing the activity rates of Neuron 4 and 5\footnote{The activity rates of Neuron 4, 5 do not depend on $\alpha$ because $\mathbf{b}_0$ does not contain interactions that relate Neuron 1-3 with Neuron 4, 5. If there are non-zero interactions between any pair from Neuron 1-3 and Neuron 4, 5 in $\mathbf{b}_0$, the activity rates of Neuron 4, 5 increase with the increased rates of Neuron 1-3.}. Furthermore, the response functions of the three neurons shift toward left when the background activity rates of all neurons is increased by \textit{decreasing} the internal component $\beta$ (Fig.~\ref{fig:gain_modulation}b dashed lines). Thus Neuron 1-3 increase sensitivity to stimulus component $\alpha$. This type of modulation is called input-gain control. For example, if $\alpha$ is a logarithmic function of contrast $s$ of visual stimulation presented to an animal while recording visual neurons ($\alpha(s)=\log s$), increasing the modulation (decreasing $\beta$) makes neurons respond to multiplicatively smaller stimulus contrast. This models the contrast gain-control observed in visual pathways \cite{Sakmann1969,Ohzawa1985,Reynolds2000,MartnezTrujillo2002}. Other types of nonlinearity in the input-output relation can be constructed, depending on the nonlinearity in $\alpha(s)$. 

Figure \ref{fig:gain_modulation}c displays a relation of the stimulus component $\alpha$ with the stimulus-related activity $X$ at different internal component $\beta$. Similarly to the activity rates (Fig.~\ref{fig:gain_modulation}b), the stimulus-related activity $X$ is augmented if the internal component $\beta$ is decreased. This nonlinear interaction between $\alpha$ and $\beta$ is caused by the neurons that belong to both stimulus-related and internal activities. In this example, the stimulus component $\alpha$ also increases the internal activity $U$ (Fig.~\ref{fig:gain_modulation}d) because of increased activity rates of the shared neurons 1, 2, 3. Finally, Figure \ref{fig:gain_modulation}e displays the variance of stimulus feature $X(\mathbf{x})$ as a function of $\alpha$. It quantifies the information about the stimulus component $\alpha$, which we will discuss in the next section.

\section{The conservation of entropy, equation of state, and stimulus information} \label{sec:first_law}

\textbf{Conservation of entropy for neural dynamics.}  
The probability mass function, Eq.~\ref{Exp_model}, belongs to the exponential family distribution. The Lagrange parameters are called natural or canonical parameters. The activity patterns of neurons are modelled as a linear combination of the two features $\mathcal{H}_0(\textbf{x})$ and $X(\textbf{x})$ using the canonical parameters $(-\beta, \alpha)$ in the exponent. Expectation of the features are called the expectation parameters $U$ and $X$. Either natural or expectation parameters are sufficient to specify the probability distribution. We review dual structure of the two representations\cite{Amari2000}, and show that the relation provides the conservation law of entropy. 

Negative entropy of the neural population is computed as 
\begin{align}
-S&= \langle \log p(\textbf{x}) \rangle \nonumber \\
    &=- \beta \langle \mathcal{H}_0(\textbf{x}) \rangle + \alpha \langle X(\textbf{x}) \rangle - \psi(\beta,\alpha) \nonumber \\
    &= - U \beta +  X \alpha - \psi(\beta,\alpha).
\label{entropy}
\end{align}
Since the log-partition function of Eq.~\ref{Exp_model} is a cumulant generating function, $U$ and $X$ are related to the derivatives of $\psi(\beta,\alpha)$ as
\begin{align}
\frac{\partial \psi(\beta,\alpha)}{\partial \beta} 
    &= -\langle \mathcal{H}_0(\mathbf{x}) \rangle = -U,  \label{dpsi_dbeta}\\
\frac{\partial \psi(\beta,\alpha)}{\partial \alpha}
    &= \langle X(\mathbf{x}) \rangle = X. \label{dpsi_dbetaf}
\end{align}

Eqs.~\ref{entropy}, \ref{dpsi_dbeta} and \ref{dpsi_dbetaf} form a Legendre transformation from $\psi(\beta, \alpha)$ to $-S(U, X)$. The inverse Legendre transformation is constructed using Eq.~\ref{entropy} as well:  $\psi(\beta, \alpha) = -\beta U  + \alpha X   -(- S(U,X))$. Thus dually to Eqs.~\ref{dpsi_dbeta} and \ref{dpsi_dbetaf}, the natural parameters are obtained as derivatives of the entropy with respect to the expectation parameters,
\begin{align}
\left( \frac{\partial S}{\partial U} \right)_{X}
    &= \beta, \\
\left( \frac{\partial S}{\partial X} \right)_{U}
    &= -\alpha. \label{dSdX}
\end{align}
The natural parameters represent sensitivities of the entropy to the independent variables $U$ and $X$. From these results, the total derivative of $S(U,X)$ is written as
\begin{align}
dS &= \left( \frac{\partial S}{\partial U} \right)_{X} dU 
+ \left( \frac{\partial S}{\partial X} \right)_{U} dX \nonumber \\
&= \beta dU - \alpha dX.
\label{first_law_dS_0}
\end{align}
This explains a change of neurons' entropy by changes in the internal and stimulus-related activities. We denote an entropy change caused by the internal activity as $dS^{\rm{int}} \equiv \beta dU$, and an entropy change caused by the extrinsic stimulus as $dS^{\rm{ext}} \equiv \alpha dX$, respectively. Then Eq.~\ref{first_law_dS_0} is written as
\begin{align}
dS = dS^{\rm{int}} - dS^{\rm{ext}} 
\label{first_law_dS}
\end{align}
We remark that $dS$ is an infinitesimal difference of entropies at two close states, and its integral does not depend on a specific transition between the two states. In contrast, $dS^{\rm{int}}$ and $dS^{\rm{ext}}$ represent production of entropy separately by the internal and stimulus-related activities, and their integrals depend on the specific paths. Eq.~\ref{first_law_dS} constitutes the conservation of entropy for neural dynamics. We stress that although it is the first law of thermodynamics, the neurons considered here interact with an environment differently from conventional thermodynamic systems\footnote{We obtain $dU = TdS - f dX $, using $\beta \equiv 1/T$ and $\alpha \equiv \beta f$ in Eq.~\ref{first_law_dS_0}. In this form, the expectation parameter $U$ is a function of $(S,X)$. According to the conventions of thermodynamics, we may call $U$ internal energy, $T$ temperature of the system, and $f$ force applied to neurons by a stimulus. It is possible to describe the evoked activity of a neural population using these standard terms of thermodynamics. However, this introduces the concepts of work and heat.}. While internal energy of the conventional systems is indirectly controlled via work and heat, we consider that the internal activity of neurons is controlled directly by the organism's internal mechanism. Thus we use $dS^{\rm{int}}$ and $dS^{\rm{ext}}$, rather than the work and heat, as quantities that neurons exchange with an environment.

\textbf{Equation of state for a neural population.}  
Eq.~\ref{dpsi_dbetaf} is an equation of the state for a neural population, which we rewrite here as 
\begin{align}
X(\beta,\alpha) = \frac{\partial \psi(\beta,\alpha)}{\partial \alpha}.
\label{eq:eq_state}
\end{align}
Through the log-partition function $\psi$, this equation relates state variables, $\beta$, $\alpha$, and $X$, similarly to e.g., the classical ideal gas law that relates temperature, pressure, and volume. Figure \ref{fig:gain_modulation}c displayed the equation of state. We note that $\psi$ is related to the Gibbs free energy (see Appendix). Furthermore, without loss of generality, we can assume that the ground state of the features is zero: $\mathcal{H}_0(\mathbf 0) =  X(\mathbf 0)= 0$, where $\mathbf{x} = \mathbf 0$ denotes the simultaneous silence of all neurons. We then obtain $p(\mathbf 0) = e^{-\psi}$, namely
\begin{align}
-\psi(\beta,\alpha) = \log p(\mathbf 0). 
\end{align}
Thus $-\psi(\beta,\alpha)$ is a logarithm of the simultaneous silence probability\footnote{Importantly, $- \psi$ is a logarithm of the simultaneous silence probability predicted by the model, Eq.~\ref{Exp_model}. The observed probability of the simultaneous silence could be different from the prediction if the model is inaccurate. For example, an Ising model can be inaccurate, and it was shown that neural higher-order interactions significantly contribute to increasing the silence probability\cite{Ohiorhenuan2010,Shimazaki2015}.}. Since $d( \log p(\textbf{0}) ) =d p(\textbf{0}) / p(\textbf{0})$, $-d \psi$ gives a fractional increase of the simultaneous silence probability of the neurons. Accordingly Eq.~\ref{eq:eq_state} states that the stimulus-related activity $X$ equals to the fractional decrease of the simultaneous silence probability by a small change of $\alpha$, given $\beta$.



\textbf{Information about stimulus.}  
The Fisher information $J(\alpha)$ provides the accuracy of estimating a small change in the stimulus component $\alpha$ by an optimal decoder. More specifically, the inverse of the Fisher information provides a lower bound of variance of an unbiased estimator for $\alpha$ from a sample. For the exponential family distribution, it is given as the second order derivative of the log-partition function with respect to $\alpha$, which is also the variance of stimulus feature $X(\mathbf{x})$: 
\begin{align}
J(\alpha) 
&\equiv \left\langle \left( \frac{\partial \log p(\mathbf{x})} {\partial \alpha} \right)^2 \right\rangle =
    \frac{\partial^2 \psi(\beta,\alpha)}{\partial \alpha^2}  \nonumber \\
    &=   \frac{\partial X}{\partial \alpha} 
    = \langle X(\mathbf{x})^2 \rangle - \langle X(\mathbf{x}) \rangle^2.
\label{J_alpha}
\end{align}
The first equality in the second line of Eq.~\ref{J_alpha} is obtained using the first order derivative of $\psi$, namely the equation of state (Eq.~\ref{eq:eq_state}). The second equality in Eq.~\ref{J_alpha} represents the fluctuation-dissipation relation of the stimulus feature. The equalities show that the Fisher information can be computed in three different manners given that the internal component $\beta$ is fixed: (i) the second derivative of $\psi$ with respect to $\alpha$ using the simultaneous silence probability, (ii) the derivative of $X$ with respect to $\alpha$ using the equation of state, or (iii) the variance of the stimulus feature.

The Fisher information computed at two fixed internal components was shown in Fig.~\ref{fig:gain_modulation}e. The stimulus component $\alpha$ becomes relatively dominant in characterising the neural activity if the internal component $\beta$ decreases. This results in the larger Fisher information $J(\alpha)$ for the smaller internal component $\beta$ at given $\alpha$. If the stimulus condition $s$ controlls the stimulus component as $\alpha(s)$, and it is not related to $\beta$, the information about $s$ is given as $\frac{\partial \alpha(s)}{\partial s} J(\alpha) \frac{\partial \alpha(s)}{\partial s}$. 

\section{Information-theoretic cycles by a neural population} \label{sec:info_engine}
We now introduce neural dynamics that models dynamical gain-modulation performed by an internal mechanism while neurons are processing stimulus. Since there are neurons that belong to both stimulus-related and internal activities, the internal mechanism not only changes the internal activity but also the stimulus-related activity, which realises the modulation. From an information-theoretic point of view, this process converts entropy generated by the internal mechanism to entropy associated with stimulus-related activity after one cycle of the neural response is completed. To explain this in detail, we first provide an intuitive example of delayed gain-modulation using a dynamical model, and then provide an ideal cycle that efficiently enhance stimulus information. Using the latter model, we explain why the process works similarly to a heat engine, and show how to quantify efficiency of the gain-modulation performed by the internal mechanism.

\begin{figure}[t]
\begin{center}
\includegraphics[width=1\textwidth]{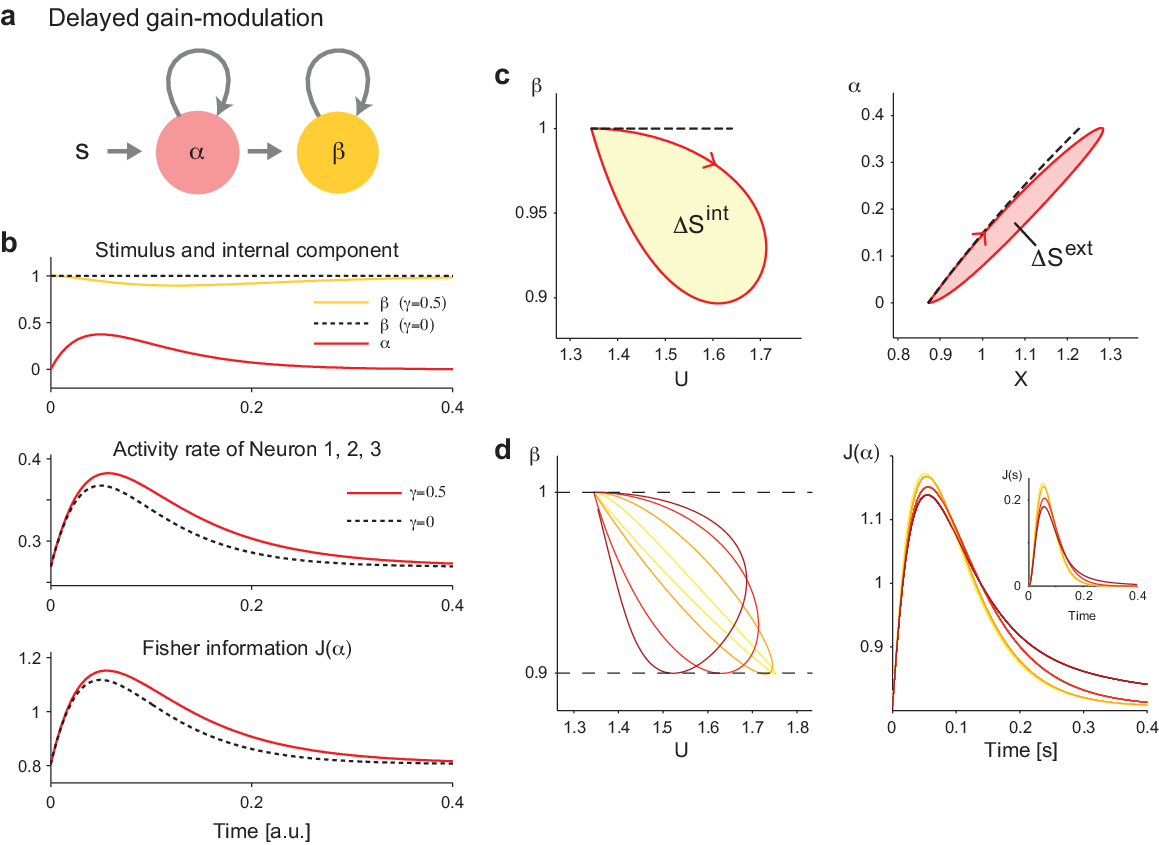}
\end{center}
\caption{
The delayed gain-modulation by internal activity. The parameters of the maximum entropy model ($N=5$) follow those in Fig.~\ref{fig:gain_modulation}. (a) An illustration of delayed gain-modulation described in Eqs.~\ref{delayed_gain1} and \ref{delayed_gain2}. The stimulus increases the stimulus component $\alpha$ that activates Neuron 1, 2, and 3. Subsequently, the internal component $\beta$ is increased, which increases the background activity of all 5 neurons. We assume a slower time constant for the gain-modulation than the stimulus activation ($\tau_{\beta}=0.1$ and $\tau_{\alpha}=0.05$). (b) Top: Dynamics of the stimulus and internal components (solid lines, $\gamma = 0.5$). The internal component $\beta$ without the delayed gain-modulation ($\gamma=0$) is shown by a dashed black line. Middle: Activity rates [a.u.] of Neuron 1-3 with (solid red) and without (dashed black) the delayed gain-modulation. Bottom: The Fisher information about stimulus component $\alpha$ (Eq.~\ref{J_alpha}). (c) The $X$-$\alpha$ (Left) and $U$-$\beta$ (Right) phase diagrams. A red solid cycle represents dynamics when the delayed gain-modulation is applied ($\gamma=0.5$). The dashed line is a trajectory when the delayed gain-modulation is not applied to the population ($\gamma=0$). (d) Left: The $U$-$\beta$ phase diagrams of neural dynamics with different combinations of $\tau_{\beta}$ and $\gamma$ that achieve the same level of the maximum modulation (the minimum value of $\beta=0.9$).  Right: The Fisher information about the stimulus component $\alpha$ for different cycles. The colour code is the same as in the Left panel.  The inset shows the Fisher information about the stimulus intensity $s$ (Eq.~\ref{Fisher_s}).
}
\label{fig:delayed_gain}
\end{figure}

\textbf{An example of delayed gain-modulation.}  We first consider a simple dynamical model of delayed gain-modulation. We use the feature vector, $\mathbf{b}_0$ and $\mathbf{b}_1$ based on those described in Fig.~\ref{fig:gain_modulation}. In this model, neurons are activated by a stimulus input, which subsequently increases modulation by an internal mechanism (Fig.~\ref{fig:delayed_gain}a). Such a process can be modelled through dynamics of the controlling parameters given by,
\begin{align}
& \tau_{\alpha}^2  \dot \alpha(t) = - \tau_{\alpha} \alpha(t) + s \, e^{-t/\tau_{\alpha}} \label{delayed_gain1} \\
& \tau_{\beta} \dot \beta(t)   		= - \beta(t) + \beta_0 -  \gamma \alpha(t) \label{delayed_gain2}
\end{align}
for $t \geq 0$. Here $s$ is intensity of an input stimulus. Neurons are initially at a spontaneous state: $\alpha(0)=0$ and $\beta(0) =  \beta_0 = 1$. The top panel of Figure~\ref{fig:delayed_gain}b displays the dynamics of $\alpha(t)$ and $\beta(t)$. The population activity is sampled from the maximum entropy model with these dynamical parameters. Here we consider a continuous-time representation of the maximum entropy model\footnote{Under the assumption that rates of synchronous spike events scale with $\mathcal{O}(\Delta^k)$, where $\Delta$ is a bin size of discretisation and $k$ is the number of synchronous neurons, Kass \emph{et al.} \cite{Kass2011} proved that it is possible to construct a continuous-time limit ($\Delta \rightarrow 0$) of the maximum entropy model that takes the synchronous events into account. Here we follow their result to consider the continuous-time representation.}\cite{Kass2011,Kelly2012}. The activity rates of neurons are increased by the delayed gain-modulation (solid lines in Fig.~\ref{fig:delayed_gain}b Middle) from those obtained without the modulation ($\gamma=0$; dashed lines). Accordingly, the information about the stimulus component $\alpha$ contained in the population activity as quantified by the Fisher information (Eq.~\ref{J_alpha}) increases and lasts longer by the delayed gain-modulation (Fig.~\ref{fig:delayed_gain}b Bottom). Note that in this example, the information about the stimulus strength $s$ is carried in both $\beta(t)$ and $\alpha(t)$ as time passes. The result obtained from the Fisher information about $s$ using both $\beta(t)$ and $\alpha(t)$ is qualitatively the same as the result of the Fisher information about $\alpha$ (not shown) \footnote{When $\alpha$ and $\beta$ are both dependent on the stimulus, the Fisher information about $s$ is given as 
\begin{align}
J(s) = \frac{ \partial \boldsymbol{\theta}(s)^T } {\partial s}  \mathbf{J} \frac{ \partial \boldsymbol{\theta}(s) } {\partial s},
\label{Fisher_s}
\end{align}
where $\boldsymbol{\theta}(s) \equiv (-\beta, \alpha)^T$ and $\mathbf{J}$ is a Fisher information matrix given by Eq.~\ref{Fisher_matrix}, which will be discussed in the later section. We computed Eq.~\ref{Fisher_s} using analytical solutions of the dynamical equations given as 
$\alpha(t) = \frac{s t}{\tau_{\alpha}} e^{-t/ \tau_{\alpha}}$ and $\beta(t) = 1 - \frac{s g}{\tau_{\beta}-\tau_{\alpha}} \left\{ \frac{\tau_{\alpha} \tau_{\beta}}{\tau_{\beta}-\tau_{\alpha}} ( e^{-t/ \tau_{\beta}}-e^{-t/\tau_{\alpha}} ) - t e^{- t/\tau_{\alpha}} \right\}$.}.

The $U$-$\beta$ phase diagram (Fig.~\ref{fig:delayed_gain}c Left) shows that dynamics without the gain-modulation is represented as a line because $\beta$ is constant. In contrast, dynamics with the gain-modulation forms a cycle because weaker and then stronger modulation (larger and then smaller $\beta$) is applied to neurons when the internal activity $U$ increases and then decreases, respectively. Similarly, the dynamics forms a cycle in the $X$-$\alpha$ plane (Fig.~\ref{fig:delayed_gain}c Right) if the stimulus activity $X$ is augmented by the delayed gain-modulation. By applying the conservation law for entropy (Eq.~\ref{first_law_dS}) to the cycle, we obtain 
\begin{align}
0 = \oint \beta dU - \oint \alpha dX.
\label{first_law_cycle}
\end{align}
Here $\oint \beta dU \equiv \Delta S^{\rm{int}}$ is entropy produced by the internal activity during the cycle due to the delayed gain-modulation, and $\oint \alpha dX \equiv \Delta S^{\rm{ext}}$ is entropy produced by the activity related to extrinsic stimulus condtions. These are the areas within the circles in the phase diagrams. Eq.~\ref{first_law_cycle} states that the two cycles have the same area ($\Delta S^{\rm{int}} = \Delta S^{\rm{ext}}$). 

The left panel in Figure \ref{fig:delayed_gain}d displays the $U$-$\beta$ phase diagram for dynamics with given maximum strength of modulation (the minimum value of $\beta$). Among these cycles, larger cycles retain the information about the stimulus component $\alpha$ for a longer time period (Fig.~\ref{fig:delayed_gain}d Right). The same conclusion is made from the Fisher information about $s$ (Fig.~\ref{fig:delayed_gain}d an inset in Right panel). The larger cycles were made because the modulation was only weakly applied to neurons when the internal activity $U$ increased, then the strong modulation was applied when $U$ decreased. Such modulation is considered to be efficient because it allows neurons to retain the stimulus information for a longer time period by using the slow time-scale of $\beta$ without excessively increasing activity rates of neurons at its initial rise. In the next section, we introduce the largest cycle that maximises the entropy produced by the gain modulation when the maximum strength of the modulation is given. Using this cycle, we explain how the cycle works analogously to a heat engine, and define efficiency of the cycle to retain the stimulus information. 

\begin{figure}[t]
\begin{center}
\includegraphics[width=1\textwidth]{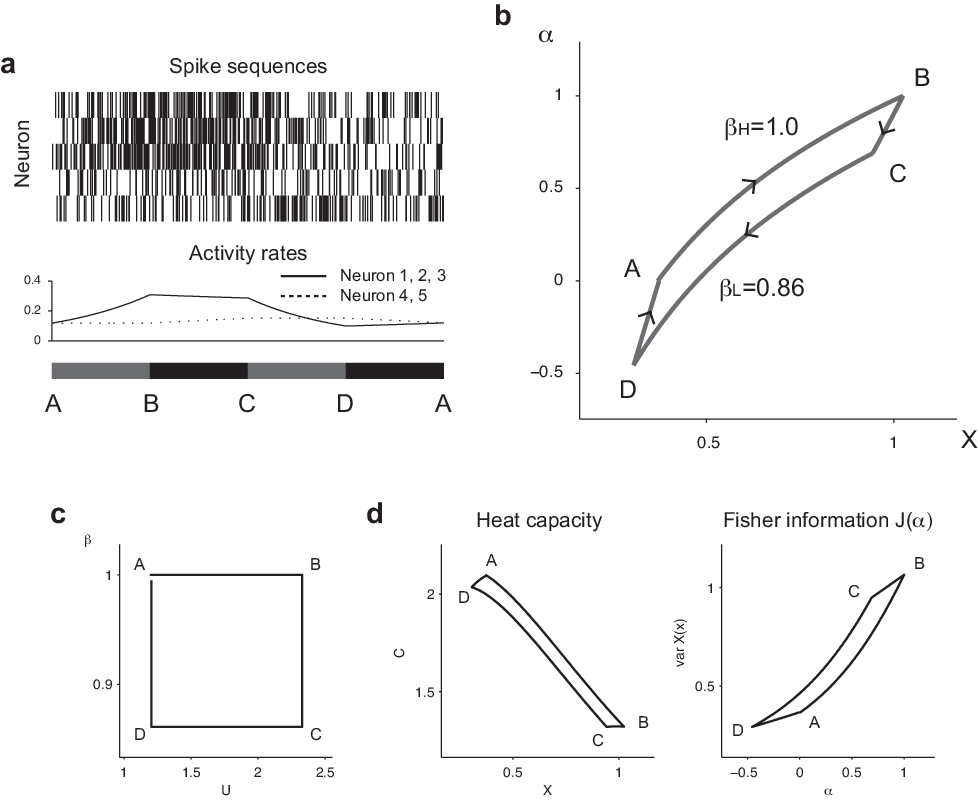}
\end{center}
\caption{The efficient circle by a neural population ($N=5$). The parameters of the maximum entropy model follow those in Fig.~\ref{fig:gain_modulation}. The cycle starts from the state A at which $\beta =\beta_H=1$ and $\alpha=0$. See the main text for details of the steps. The efficiency of this cycle is $0.14$. (a) Top: Spike raster plots during the cycle. Middle: Activity rates of neurons. Bottom: The cycle steps. (b) The $X$-$\alpha$ phase diagram. (c) The $U$-$\beta$ phase diagram. (d) Left: $X$ v.s. heat capacity. The heat capacity is defined as $C = \langle h^2 \rangle - \langle h \rangle ^2$, where $h=-\log p(\mathbf{x}) $ is information content. Right: Fisher information about the stimulus component $\alpha$. 
}
\label{information_cycle}
\end{figure}

\textbf{The efficient cycle by a neural population.}  
The largest cycle is made if the modulation is not applied when the internal activity $U$ increases, then applied when $U$ decreases. Figure \ref{information_cycle} displays a cycle of hypothetical neural dynamics that maximises the entropy production when the ranges of the internal component and activity are given. The model parameters follow those in Fig.~\ref{fig:gain_modulation}. This cycle is composed of four steps. The process starts at the state A at which neurons exhibit spontaneous activity ($\beta = \beta_H =1$, $\alpha=0$). Figure \ref{information_cycle}a displays a sample response of the neural population to a stimulus change. Figure \ref{information_cycle}b and c display the $X$-$\alpha$ and $U$-$\beta$ phase diagrams of the cycle. Heat capacity of the neural population and the Fisher information about $\alpha$ are shown in Fig.~\ref{information_cycle}d. Details of the cycle steps are now described as follows.

\vspace{.5em} 

\begin{enumerate} 
  \item[A$\rightarrow$B]  \textbf{Increased stimulus response} The stimulus-related activity $X$ is increased by increasing the stimulus component $\alpha$ while the internal component is fixed at $\beta = \beta_H$. In this process the internal activity $U$ also increases. 
    
  \item[B$\rightarrow$C]  \textbf{Internal computation} An internal mechanism decreases the internal component $\beta$ while keeping the internal activity ($dU=0$). In this process the stimulus-related activity $X$ decreases. The process ends at $\beta = \beta_L$. 
  
  \item[C$\rightarrow$D]  \textbf{Decreased stimulus response} The stimulus-related activity $X$ is decreased by decreasing the stimulus component $\alpha$ while the internal component is fixed at $\beta = \beta_L$. In this process the internal activity $U$ also decreases. 

  \item[D$\rightarrow$A]  \textbf{Internal computation} An internal mechanism increases the internal component $\beta$ while keeping the internal activity ($dU=0$). In this process the stimulus-related activity $X$ increases. The process ends at $\beta \equiv \beta_H$. 
  
\end{enumerate}

The processes B$\rightarrow$C and D$\rightarrow$A represent additional computation performed by an internal neural mechanism on the neurons' stimulus information processing. It is applied after the initial increase of stimulus-related activity during A$\rightarrow$B, therefore manifests delayed modulation. Without these processes, the neural dynamics is represented as a line in the phase diagrams. The Fisher information about $\alpha$ also increases during the process between C and D (Fig.~\ref{information_cycle}d right panel). We reiterate that the Fisher information quantifies the accuracy of estimating a small change in $\alpha$ by an optimal decoder. Thus operating along the path between C and D is more advantageous than the path between A and B for downstream neurons if their goal is to detect a change in the stimulus-related activity of the upstream neurons that is not explained by the internal activity. 

\textbf{Interpretation as an information-theoretic cycle.} We start our analysis on the cycle by examining how much entropy is generated by the internal and stimulus-related activities at each step. First, we denote by $\Delta S^{\rm{int}}_{\rm{AB}}$ and $\Delta S^{\rm{int}}_{\rm{CD}}$ the entropy changes caused by the internal activity during the process A$\rightarrow$B and C$\rightarrow$D, respectively. Since the internal component $\beta$ is fixed at $\beta_H$ during the process A$\rightarrow$B, we obtain $\Delta S^{\rm{int}}_{\rm{AB}} = \beta_H \Delta U$, where $\Delta U$ is a change of the internal activity (see Fig.~\ref{information_cycle}c). This change in the internal activity is positive ($\Delta U>0$). Since the internal activity does not change during B$\rightarrow$C and D$\rightarrow$A, a change of the internal activity during C$\rightarrow$D is given by $-\Delta U$ (Note that the internal activity is a state variable). We obtain $\Delta S^{\rm{int}}_{\rm{CD}}= - \beta_L \Delta U$ for the process during C$\rightarrow$D. The total entropy change caused by the internal activity during the cycle is given as $\Delta S^{\rm{int}}_{\rm{AB}} + \Delta S^{\rm{int}}_{\rm{CD}} = (\beta_H - \beta_L) \Delta U$, which is positive because $\beta_H > \beta_L$ and $\Delta U>0$. Thus the internal activity increases the entropy of neurons during the cycle. Second, we denote by $\Delta S^{\rm{ext}}$ the total entropy change caused by the stimulus-related activity during the cycle. According to the conservation law (Eq.~\ref{first_law_dS}) applied to this cycle, we obtain 
\begin{align}
0 = \Delta S^{\rm{int}}_{\rm{AB}} + \Delta S^{\rm{int}}_{\rm{CD}} - \Delta S^{\rm{ext}}.
\end{align} 
Note that the sign of $\Delta S^{\rm{ext}}=  \Delta S^{\rm{int}}_{\rm{AB}} + \Delta S^{\rm{int}}_{\rm{CD}}$ is positive. Hence the stimulus-related activity decreases the entropy of neurons during the cycle. 

\begin{figure}[t]
\begin{center}
\includegraphics[width=.5\textwidth]{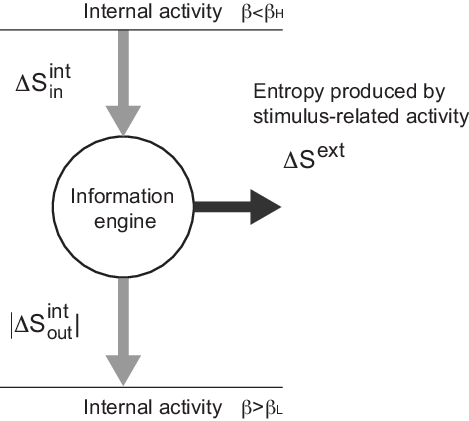}
\end{center}
\caption{An information-theoretic cycle by a neural population. 
}
\label{information_cycle_schematics}
\end{figure}

This cycle belongs to the following cycle that is analogous to a heat engine (Fig.~\ref{information_cycle_schematics}). In this paragraph, we temporarily use \textit{receive entropy} and \textit{emit entropy} to express the positive and negative path-dependent entropy changes caused by the internal or stimulus-related activity in order to facilitate comparison with a heat engine\footnote{Here we use \textit{entropy} synonymously with heat in thermodynamics to facilitate the comparison with a heat engine. However this is not an accurate description because the entropy is a state variable.}. In this cycle, neurons receive \textit{entropy} as internal activity from an environment ($\Delta S^{\rm{int}}_{\rm{in}}>0$) and emit \textit{entropy} to the environment ($\Delta S^{\rm{int}}_{\rm{out}}<0$). The received \textit{entropy} as the internal activity is larger than the emitted \textit{entropy} ($\Delta S^{\rm{int}}_{\rm{in}} + \Delta S^{\rm{int}}_{\rm{out}}>0$). The surplus \textit{entropy} is emitted to the environment in the form of the stimulus-related activity ($-\Delta S^{\rm{ext}}<0$). Thus we may regard the cycle as the process that produces stimulus-related entropy using entropy supplied by the internal mechanism. We hereafter denote this cycle as an information-theoretic cycle, or engine. The cycle in Fig.~\ref{fig:delayed_gain} is also regarded as an information-theoretic cycle by separating the process at which the internal activity is maximised. The conservation law prohibits a perpetual information-theoretic cycle that can indefinitely produce the stimulus-related entropy without entropy production by the internal mechanism\footnote{This is synonymous with the statement that the first law prohibits a perpetual motion machine of the first kind, a machine that can work indefinitely without receiving heat.}. 

\textbf{Efficiency of a cycle.} 
As we discussed for the example dynamics in Fig.~\ref{fig:delayed_gain}, we may consider that the modulation is efficient if it helps neurons to retain stimulus information without excessively increasing the internal and stimulus-related activities during the initial response. Such a process was achieved when gain modulation was only weakly applied to neurons when the internal activity $U$ increased, then strong gain modulation was applied when $U$ decreased. We can formally assess this type of efficiency by defining entropic efficiency, similarly to thermal efficiency of a heat engine. It is given by a ratio of the entropy change caused by the stimulus-related activity as opposed to the entropy change gained by the internal activity as:  
\begin{align}
\eta &\equiv \frac{ \Delta S^{\rm{ext}} } {\Delta S^{\rm{int}}_{\rm{in}} }  
	=   1 - \frac{| \Delta S^{\rm{int}}_{\rm{out}} |} {\Delta S^{\rm{int}}_{\rm{in}} }.  
	\label{efficiency} 
\end{align}
For the proposed information-theoretic cycle in Fig.~\ref{information_cycle}, it is computed as
\begin{align}
	\eta_{e} = 1 - \frac{| \Delta S^{\rm{int}}_{\rm{CD}} |} {\Delta S^{\rm{int}}_{\rm{AB}} } =  1 - \frac{\beta_L} {\beta_H }, 
\end{align}
which is a function of the internal components, $\beta_H$ and $\beta_L$. This cycle is the most efficient in terms of the entropic efficiency defined by Eq.~\ref{efficiency} when the highest and lowest internal components and activities are given. The square cycle in the $U$-$\beta$ phase diagram (Fig.~\ref{information_cycle}c) already suggests this claim, and we can formally prove this by comparing the information-theoretic cycle with an arbitrary cycle $\mathcal{C}$ whose internal component $\beta$ satisfies $\beta_L \leq \beta \leq \beta_H$\footnote{Let us consider the efficiency $\eta$ achieved by an arbitrary cycle $\mathcal{C}$ during which the internal component $\beta$ satisfies $\beta_L \leq \beta \leq \beta_H$. Let the minimum and maximum internal activity in the cycle be $U_{\rm{min}}$ and $U_{\rm{max}}$. We decompose $\mathcal{C}$ into the path $\mathcal{C}_1$ from $U_{\rm{min}}$ to $U_{\rm{max}}$ and the path $\mathcal{C}_2$ from $U_{\rm{max}}$ to $U_{\rm{min}}$ during which the internal component is given as $\beta_1(U)$ and $\beta_2(U)$, respectively. Because the cycle acts as an engine, we expect $\beta_1(U)>\beta_2(U)$. The entropy changes produced by the internal activity during the path $C_i$ ($i=1,2$) is computed as $\Delta S^{\rm{int}}_{\mathcal{C}_1} = \int_{U_{\rm{min}}}^{U_{\rm{max}}} \beta_1(U) \, dU \leq \beta_H \int_{U_{\rm{min}}}^{U_{\rm{max}}}  \, dU = \beta_H (U_{\rm{max}}-U_{\rm{min}})$ and  $| \Delta S^{\rm{int}}_{\mathcal{C}_2} |= |\int_{U_{\rm{max}}}^{U_{\rm{min}}} \beta_2(U) \, dU| \geq |\beta_L \int_{U_{\rm{max}}}^{U_{\rm{min}}} \, dU| = \beta_L (U_{\rm{max}}-U_{\rm{min}})$. Hence we obtain $| \Delta S^{\rm{int}}_{\mathcal{C}_2} | / \Delta S^{\rm{int}}_{\mathcal{C}_1} \geq \beta_L / \beta_H  $, or $\eta \leq \eta_e$.}. Thus the proposed cycle bounds efficiency of the additional computation made by the delayed gain-modulation mechanism. We call the proposed cycle in Fig.~\ref{information_cycle}, the ideal information-theoretic cycle. Note that this cycle is similar to, but different from the Carnot cycle \cite{Carnot1824} that can be realised by replacing the processes B$\rightarrow$C and D$\rightarrow$A with adiabatic processes. The Carnot cycle achieves the highest \textit{thermal} efficiency.

\textbf{Geometric interpretation.} 
Finally, to consider conditions for the information-theoretic cycle, we introduce geometric interpretation of the cycle. 
Let us denote the internal and stimulus components as $\boldsymbol{\theta} = (-\beta, \alpha)^T$. In addition, we represent the expected internal and stimulus features by $\boldsymbol{\eta} = (U, X)^T$. The parameters $\boldsymbol{\theta}$ and $\boldsymbol{\eta}$ form dually flat affine coordinates, and are called $\theta$ and $\eta$-coordinates in information geometry\cite{Amari2000}. For the ideal information-theoretic cycle, we indicate the parameters at A, B, C, and D using a subscript of $\boldsymbol{\theta}$ or $\boldsymbol{\eta}$. For example the parameters at A are $\boldsymbol{\theta}_A$ and $\boldsymbol{\eta}_A$. 

The first process A$\rightarrow$B of the ideal information-theoretic cycle is a straight line (geodesic) between $\boldsymbol{\theta}_{\rm{A}}$ and $\boldsymbol{\theta}_{\rm{B}}$ in the curved space of $\theta$-coordinates. It is called e-geodesic. In addition, the internal component $\beta$ is fixed while the stimulus component decreases, therefore the e-geodesic is a vertical line in the $\theta$-coordinates. The second process B$\rightarrow$C is the shortest line between $\boldsymbol{\eta}_{\rm{B}}$ and $\boldsymbol{\eta}_{\rm{C}}$ in the curved space of $\eta$-coordinates. The path is called an m-geodesic. In addition, the internal activity $U$ is fixed while the stimulus-related activity decreases, therefore the m-geodesic is a vertical line in the $\eta$-coordinates. Similarly, the process C$\rightarrow$D is an e-geodesic, and the process D$\rightarrow$A is an m-geodesic. 

The change in the internal component $\beta$ during the processes along m-geodesic manifested the internal computation in the ideal information-theoretic cycle. A small change in $\boldsymbol{\theta}$ is related to a change in $\boldsymbol{\eta}$ as $d \boldsymbol{\eta} = \mathbf{J} d \boldsymbol{\theta}$. Here $\mathbf{J}$ is the Fisher information matrix with respect to $\boldsymbol{\theta}$. It is given as
\begin{align} 
\mathbf{J}
=  \left[ 
\begin{array}{cc}
\langle \mathbf{b}_0 , \mathbf{b}_0 \rangle & \langle \mathbf{b}_0 , \mathbf{b}_1 \rangle \\
\langle \mathbf{b}_1, \mathbf{b}_0 \rangle  &  \langle \mathbf{b}_1, \mathbf{b}_1 \rangle
\end{array}
\right], 
\label{Fisher_matrix}
\end{align} 
where $\langle \mathbf{b}_i , \mathbf{b}_j \rangle \equiv \mathbf{b}_i^T \mathbf{G} \mathbf{b}_j$ $(i,j=0,1)$ is an inner product of the vectors $\mathbf{b}_i$ and $\mathbf{b}_j$ with a metric given by $\mathbf{G} = \langle \mathbf{F}(\mathbf{x}) \mathbf{F}(\mathbf{x})^T \rangle - \langle \mathbf{F}(\mathbf{x}) \rangle \langle \mathbf{F}(\mathbf{x}) \rangle^T$. Note that $\langle \mathbf{b}_0 , \mathbf{b}_0 \rangle$ is equivalent to Eq.~\ref{J_alpha}. Likewise, the small change in $\boldsymbol{\eta}$ is related to the change in $\boldsymbol{\theta}$ by $d \boldsymbol{\theta} = \mathbf{J}^{-1} d \boldsymbol{\eta}$. Since the m-geodesic processes B$\rightarrow$C and D$\rightarrow$A are characterised by $d \boldsymbol{\eta} = (0,dX)^T$, the small change in $\theta$-coordinates is given as
\begin{align} 
d \boldsymbol{\theta} 
=
\left[ 
\begin{array}{c}
- \langle \mathbf{b}_0 , \mathbf{b}_1 \rangle  \\
\langle \mathbf{b}_0, \mathbf{b}_0 \rangle 
\end{array}
\right] |\mathbf{J}|^{-1} dX, 
\end{align}
Conversely, the internal mechanism needs to change the internal and stimulus component according to the above gradient in order to accomplish the most efficient cycle. Thus if the internal mechanism can not access the stimulus component $\alpha$, the ideal information-theoretic cycle is not realised. Further, if $\langle \mathbf{b}_0 , \mathbf{b}_1 \rangle = 0$, the internal component $\beta$ is not allowed to change, which however means that the entire process does not form a cycle. Therefore we impose $\langle \mathbf{b}_0 , \mathbf{b}_1 \rangle \neq 0$. This equation indicates that the modulation by an internal mechanism is achieved through the activity features shared by the two components. Accordingly, this condition is violated if neurons participate in the stimulus-related activity and neurons subject to the internal modulation do not overlap (namely if neurons that appear in the features corresponding to non-zero elements of $\mathbf{b}_0$ are separable from those of $\mathbf{b}_1$). In general, in order to make a change of the internal component $\beta$ influence the stimulus-related activity $X$, therefore controls stimulus information, one requires $\langle \mathbf{b}_0 , \mathbf{b}_1 \rangle \neq 0$ because $dX = - \langle \mathbf{b}_1, \mathbf{b}_0 \rangle d\beta + \langle \mathbf{b}_1 , \mathbf{b}_1 \rangle d\alpha$ from $d \boldsymbol{\eta} = \mathbf{J} d \boldsymbol{\theta}$.

\section{Discussion}
In this study, we provided hypothetical neural dynamics that efficiently encodes stimulus information with the aid of delayed gain-modulation by an internal mechanism, and demonstrated that the dynamics forms an information-theoretic cycle that acts similarly to a heat engine. This view provided us to quantify the efficiency of the gain-modulation in retaining the stimulus information. The ideal information-theoretic cycle introduced here bounded the entropic efficiency. 

As an extension of a logistic activation function of a single neuron to multinomial outputs, the maximum entropy model explains probabilities of activity patterns by a softmax function of the features, therefore allows nonlinear interaction of the inputs (here $\beta$ and $\alpha$) in producing the stimulus-related activity $X$ (Fig.~\ref{fig:gain_modulation}). This interaction was caused by shared activity features in $\mathbf{b}_1$ and $\mathbf{b}_0$. The gain modulation more effectively changes the stimulus-related activity if the features of the stimulus-related and internal activities resemble (i.e., $\langle \mathbf{b}_1, \mathbf{b}_0 \rangle $ is close to $1$), which may have implications in similarity between evoked and spontaneous activities \cite{Kenet2003} that can be acquired during development \cite{Berkes2011}. 

The model's statistical structure common to thermodynamics (the Legendre transformation; see Appendix) allowed us to construct the first law for neural dynamics (Eq.~\ref{first_law_dS}), the equation of state (Eq.~\ref{eq:eq_state}), fluctuation-dissipation relation (Eq.~\ref{J_alpha}), and neural dynamics similar to a thermodynamic cycle (Figs.~\ref{fig:delayed_gain} and \ref{information_cycle}) although we emphasised the differences from conventional thermodynamics in terms of the controllable quantities. The dynamics forms a cycle if the gain modulation is applied after the initial increase of the stimulus-related activity. This scenario is expected when the stimulus response is modulated by a feedback mechanism of recurrent networks \cite{Salinas1996,Spratling2004,Sutherland2009}, and is associated with short-term memory of the stimulus \cite{Salinas1996,Salinas2001}. Consistently with the idea of efficient stimulus-encoding by a cycle, effect of attentional modulation on neural response typically appears several hundred milliseconds after stimulus onset (later than the onset of the stimulus response) \cite{Motter1993,Luck1997,McAdams1999,Seidemann1999,Reynolds2000,Ghose2002} although the temporal profile can be altered by task design \cite{Luck1997,Ghose2002}. 

To apply the theory to empirical data, the internal and stimulus feature need to be specified. Since even spontaneous neural activity is known to exhibit ongoing dynamics\cite{Kenet2003}, estimation of these features is nontrivial. The optimal sequential Bayesian algorithms have been proposed to smoothly estimate the parameters of the neural population model when they vary in time \cite{Shimazaki2009,Shimazaki2012,Shimazaki2013}. These approaches can be used to select dominant features of spontaneous and evoked activities, and then to estimate the time-varying internal and stimulus-related components. By including multiple stimulus features in the model, the theory is expected to make quantitative predictions on competitive mechanisms of selective attention \cite{Moran1985,Motter1993,Luck1997,Reynolds1999}. The conservation law of entropy imposes competition among the stimuli given a limited entropic resource generated by the internal mechanism. 


In sum, a neural population that works as an information-theoretic engine produces entropy ascribed to stimulus-related activity out of entropy supplied by an internal mechanism. This process is expected to appear during stimulus response of neurons subject to feedback gain-modulation. It is thus hoped that quantitative assessment of the neural dynamics as an information-theoretic cycle contributes to understanding neural computation performed internally in an organism. 


\subsubsection*{Acknowledgments}
The author thanks C Donner, D Hirashima, S Koyama, and S Amari for critically reading the manuscript. 

\appendix

\section*{Appendix: Free energies of neurons} \label{free_energies}
In this appendix, we introduce free energies of a neural population. Let us first discuss the relation of state variables and free energies that appear in our analysis of the neural population with those found in conventional thermodynamics. Assume that the small change in internal activity of neurons has the following linear relations to entropy $S$, expected feature $X$, and the number of neurons $N$:
\begin{align}
dU = TdS + f dX + \mu dN. 
\label{first_law_thermo}
\end{align}
Eq.~\ref{first_law_thermo} is the first law of thermodynamics, and the parameters are temperature $T$, force $f$, and chemical potential $\mu$. 
The first law describes the internal activity as a function of $(S,X,N)$. In thermodynamics, the Helmholtz free energy $F=U-TS$ , Gibbs free energy $G = F - fX$, or enthalpy $H = U - fX$ are introduced to change the independent variables to $(T,X,N)$, $(T,f,N)$, and $(S,f,N)$, respectively. These free energies are useful to analyse isothermal or other processes in which only one of the independent variables is changed. For example, the Helmholtz free energy can be used to compute the work done by force $f$ under the isothermal condition. However, the concepts of the force and work may not be directly relevant to information-theoretic analysis of a neural population. Here we introduce the free energies that are more consistent with the framework based on entropy changes. 

The first law is alternatively written as 
\begin{align}
dS =  \beta dU - \alpha dX - \gamma dN, 
\label{first_law_thermo_dS}
\end{align}
Here we used $\beta = 1/T$, $\alpha = f/T$, and $\gamma= \mu/T$. This first law describes a small entropy change as a function of $(U,X,N)$. The parameters are defined as
\begin{align}
\beta(U,X,N) &= \left(\frac{\partial S}{\partial U} \right)_{X,N}, \\
\alpha(U,X,N) &= - \left(\frac{\partial S}{\partial X} \right)_{N,U}, \\
\gamma(U,X,N) &= - \left(\frac{\partial S}{\partial N} \right)_{U,X}.
\end{align}
We change the independent variable $U$ to $\beta$. For this goal, here we define the \textit{scaled} Helmholtz free energy $\mathcal{F}$ as
\begin{align}
\mathcal{F} = S - \beta U. 
\end{align}
Note that $\mathcal{F}=-\beta F$. It is a function that changes the independent variables from $(S,X,N)$ to $(\beta,X,N)$. This can be confirmed from the total derivative of $\mathcal{F}$: $d\mathcal{F} = dS - d(\beta U) = - U d\beta - \alpha dX - \gamma dN$. From this equation, we have
\begin{align}
U(\beta,X,N) &=  - \left(\frac{\partial \mathcal{F}}{\partial \beta} \right)_{X,N}, \\ 
\alpha(\beta,X,N)  &= - \left(\frac{\partial \mathcal{F}}{\partial X} \right)_{N,\beta}, \label{alpha_FX}\\
\gamma (\beta,X,N) &= - \left(\frac{\partial \mathcal{F}}{\partial N} \right)_{\beta,X}.
\label{Force_f}
\end{align}
The entropy change caused by the stimulus-related activity when $X$ changes from $X_1$ to $X_2$ is given by the area under the curve of $\alpha(\beta,X,N) $ in the $X\textrm{-}\alpha$ phase plane. From Eq.~\ref{alpha_FX}, if the process satisfies $d\beta=dN=0$, the entropy change is computed as reduction of the scaled Helmholtz free energy as
\begin{align}
\Delta S^{\rm{ext}}   = \int_{X_1}^{X_2}  \alpha(\beta,X,N) \, dX 
		= \mathcal{F}(\beta,X_2,N)-\mathcal{F}(\beta,X_1,N).
\label{work_F}
\end{align}


Further change of the independent variables from $(\beta,X,N)$ to $(\beta,\alpha,N)$ is done by introducing the \textit{scaled} Gibbs free energy:
\begin{align}
\mathcal{G} = \mathcal{F} + \alpha X = S - \beta U + \alpha X.
\label{Gibbs_free_energy}
\end{align}
Note that $\mathcal{G}=-\beta G$. The independent variables of the Gibbs free energy are $(\beta,\alpha,N)$ since $d\mathcal{G} = d\mathcal{F} + (d\alpha X + X d\alpha) = - U d\beta + X d\alpha - \gamma dN$. From this equation, we find 
\begin{align}
\left(\frac{\partial \mathcal{G}}{\partial \beta} \right)_{\alpha,N} &= - U(\beta,\alpha,N), \label{G_beta}\\
\left(\frac{\partial \mathcal{G}}{\partial \alpha} \right)_{\beta,N} &= X(\beta,\alpha,N). \label{G_alpha}
\end{align}
Note that the definition of the Gibbs free energy by Eq.~\ref{Gibbs_free_energy} is obtained from Eq.~\ref{entropy} if we identify $\mathcal{G} = \psi$. Accordingly, Eqs.~\ref{G_beta} and \ref{G_alpha} coincide with Eqs.~\ref{dpsi_dbeta} and \ref{dpsi_dbetaf}. 


\bibliography{exported-references}{}
\bibliographystyle{abbrv} 

\end{document}